%% file: main.tex
\documentclass[a4paper, 10pt, conference]{ieeeconf}  

\IEEEoverridecommandlockouts                              
\usepackage{amsmath}
\usepackage{amsfonts}
\usepackage{subcaption}
\usepackage{graphicx}
\usepackage{gensymb}
\usepackage{booktabs}       
\usepackage{diagbox}
\usepackage{tikz}
\usetikzlibrary{shapes, arrows, calc}
\usepackage{carshapes}
\usepackage{numprint}
\usepackage[misc]{ifsym}
\usepackage[nolist,nohyperlinks]{acronym}
\begin{acronym}
	\acro{ad}[AD]{Automated Driving}
    \acro{adas}[ADAS]{Advanced Driver-Assistance System}
	\acro{bmc}[BMC]{Bounded Model Checker}
	\acro{odd}[ODD]{Operational Design Domain}
	\acro{acc}[ACC]{Adaptive Cruise Control}
	\acro{mpc}[MPC]{Model Predictive Control}
	\acro{lka}[LKA]{Lane Keeping Assist}
	\acro{qp}[QP]{Quadratic Programming}
 	\acro{dnn}[DNN]{Deep Neural Network}
    \acro{cis}[CIS]{Controlled Invariant Set}
    \acro{smt}[SMT]{Satisfiability Modulo Theory}
    \acro{cps}[CPS]{Cyber-Physical Systems}
    \acro{ddpg}[DDPG]{Deep Deterministic Policy Gradient}
    \acro{vhc}[VHC]{Vehicle Hardware Configuration}
    \acro{method}[CoCoSaFe]{\underline{Co}mpositional \underline{Co}de-level \underline{Sa}fety verification \underline{F}ram\underline{e}work}
\end{acronym}

\title{\LARGE \bf
One Stack, Diverse Vehicles: Checking Safe Portability of Automated Driving Software
}

\author{Vladislav Nenchev
\thanks{V.~Nenchev is with the Department of Electrical and Computer Engineering, University of the Bundeswehr Munich, Werner-Heisenberg-Weg 39, 85579 Neubiberg, Germany. He was formerly with the BMW Group, Munich. {\tt\small vladislav.nenchev@unibw.de}}%
}

\begin{document}
\maketitle
\thispagestyle{empty}
\pagestyle{empty}

\input{sections/abstract.tex}
\input{sections/introduction.tex}

\input{sections/problem_statement.tex}
\input{sections/framework.tex}
\input{sections/case_study.tex}
\input{sections/conclusions.tex}

\bibliographystyle{IEEEtran}
\bibliography{IEEEabrv,references}

\end{document}

%% file: sections/abstract.tex
\begin{abstract}
Integrating an automated driving software stack into vehicles with variable configuration is challenging, especially due to different hardware characteristics. Further, to provide software updates to a vehicle fleet in the field, the functional safety of every affected configuration has to be ensured. These additional demands for dependability and the increasing hardware diversity in automated driving make rigorous automatic analysis essential. This paper addresses this challenge by using formal portability checking of adaptive cruise controller code for different vehicle configurations. Given a formal specification of the safe behavior, models of target configurations are derived, which capture relevant effects of sensors, actuators and computing platforms. A corresponding safe set is obtained and used to check if the desired behavior is achievable on all targets. In a case study, portability checking of a traditional and a neural network controller are performed automatically within minutes for each vehicle hardware configuration. The check provides feedback for necessary adaptations of the controllers, thus, allowing rapid integration and testing of software or parameter changes. 

\end{abstract}

%% file: sections/introduction.tex
\section{Introduction}\label{sec:intro}

An \ac{ad} software stack is typically developed focusing on one \ac{vhc} initially. In the later stages of development these must be ported to operate across a wide range of hardware including different sensor types, actuators, and computing platforms. Traditionally, this process has been manual, iterative and, thus, error prone. Further, an emerging business model in the automotive domain is function-on-demand, where customers initially purchase a vehicle without certain features and later upgrade the functionality as needed. This is often facilitated through over-the-air software updates, which pose additional safety challenges due to the complexity of interacting software and hardware components and the high configurability of modern vehicles.

The safety of an \ac{ad} function can be compromised by the performance of sensors, actuators, and computing devices \cite{Gordon2015}. Extensive simulation and real-world testing are typically used to ensure safe operation in a wide range of scenarios. While software testing against selected test cases has proven useful to prevent regression and supports debugging during development, it is non-exhaustive by nature. Formal methods may provide guarantees of completeness to a certain level in the context of AD \cite{Mehdipour2023}, but they are mainly based on abstract models derived manually from the software deployed in the vehicle. Automatic portability checking methods at code-level become increasingly important for managing hardware diversity and covering the demand for reliable \ac{ad} systems with a large \ac{odd} including after-sale (e.g. over-the-air) updates and feature releases.

In this paper, formal models of the vehicle and its surroundings are used to automatically check the portability of \ac{acc} software to different \ac{vhc}s. The derived models encompass common safety-relevant aspects of sensors, computing platforms, and actuators in \ac{ad}. Given the main functional safety requirement of \ac{acc} \cite{ACCISO2018} -- to maintain an appropriate safe distance from relevant front objects -- a safe set within the \ac{odd} is obtained for each \ac{vhc}. This allows using the method proposed in \cite{Nenchev2024} to check if the closed-loop execution of the \ac{acc} code operates safely for each \ac{vhc}. A case study demonstrates how the check provides essential feedback for conventionally implemented and \ac{dnn}-based \ac{acc} software for three different \ac{vhc}s. 

The proposed portability check comprehensively covers the entire \ac{odd}, but relies on a formal model that may not reflect all behaviors of the real system. Although real-world datasets provide a more accurate representation of behaviors, they may fail to encompass all relevant scenarios within the \ac{odd}. Therefore, this approach serves as a complementary method to evaluations based on real-world datasets. It supports rapid integration and testing of software or parameter variations, thus enhancing the efficiency of deploying AD functions across a diverse vehicle fleet. 

The remainder of the paper is organized as follows. After summarizing related work (Sec.\,\ref{ref:related_work}), preliminaries for ACC are provided (Sec.\,\ref{sec::problem}). Then, in Sec.\,\ref{sec:framework}, models and safe sets for diverse configurations are obtained to enable portability checking. Finally, the portability of two ACC implementations is checked for three \ac{vhc}s (Sec.\,\ref{sec:experiments}), followed by a discussion and conclusions (Sec.\,\ref{sec:conclusions}).

\section{Related work}\label{ref:related_work}

Testing is common practice for checking features and functional safety, and is also often applied to show the portability or compatibility of software (updates). However, software compatibility testing \cite{Haghi2017} is traditionally non-functional, i.e., it ensures that the software application can be executed on the hardware (possibly including a different operating system, communication network etc.). Functional portability testing checks if the software operates according to specified functional requirements that is traditionally implemented as a combination of unit, simulation and manual real-world tests. Functional testing is often complemented by an automatic search for specification counter-examples, e.g., using reinforcement learning \cite{Favrin2020}, sampling promising operation regions \cite{Chou2018}. or extracting abstract logic models from code \cite{Nenchev2021}. Control software executions can also be checked with respect to monitors, which encode formal specifications \cite{Nenchev2023}. Although these methods can identify corner cases that might otherwise be missed, they are not guaranteed to scale effectively for all automotive systems, nor do they offer portability statements for the entire \ac{odd} of an \ac{ad} function.

By using model checking \cite{Lygeros1996}, counter-example-guided searching \cite{Stursberg2004} or reachability analysis \cite{Alam2014} formal verification of traditionally designed controllers can be achieved with some completeness guarantees. Automatic proofs of closed-loop safety have also been obtained for DNN-based controllers, e.g., \cite{Dawson2023,Sun2019}. However, these methods do not operate directly with the deployed code, but employ conservative approximations or a model of the controller. 

Porting software to new configurations was addressed by deriving a hardware-specific model for formally checking portability and auto-generating low-level code for different OS kernels and drivers \cite{Gomes2024}. A case study for code-level safety verification based on set invariance was presented for ACC \cite{Nenchev2024}. In contrast to \cite{Nenchev2024}, the models employed in this work capture not only additional relevant aspects of the vehicle's basic kinematics and its environment, but also important points of variation for \ac{vhc}s such as the ego vehicle's dynamics and sensing properties. 

To the best of the author's knowledge, this paper is the first to provide an automatic portability check for AD software based on a formal model of safety-relevant configuration-specific functional behavior.

%% file: sections/problem_statement.tex
\section{Preliminaries}\label{sec::problem}

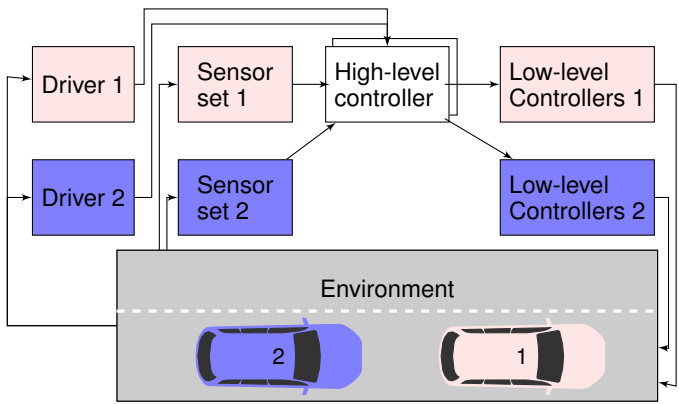
\begin{figure}[t]
 \centering
\begin{tikzpicture}[auto,>=latex', transform shape,
    every node/.style={font=\sffamily\small, node distance=0.5cm, align=left},
    block/.style={
        rectangle, draw, align=left, minimum width=1cm,
        minimum height=1cm, font=\sffamily\small
    },
    multinode/.pic={
        \node [block, \tikzpictextoptions](n) at (0,0) {\tikzpictext};
        \path [block] 
          ([shift={(.1,0)}]n.north west)
          --+(0,.1)
          -|([shift={(.1,.1)}]n.south east)
          --+(-.1,0)
        ;
    }]
    
        \node [draw, fill=red!10, rectangle, minimum width=1cm, minimum height=1cm] (driver1) {Driver 1};
    \node [draw, fill=blue!50, rectangle, below of=driver1, node distance=1.5cm, minimum width=1cm, minimum height=1cm] (driver2) {Driver 2};
    
    \node [draw, fill=red!10, rectangle, right of=driver1, node distance=2cm, minimum width=1.5cm, minimum height=1cm] (sensor1) {Sensor\\ set 1};
    \node [draw, fill=blue!50, rectangle, below of=sensor1, node distance=1.5cm, minimum width=1.5cm, minimum height=1cm] (sensor2) {Sensor\\ set 2};
    
    \node [right of=sensor1, node distance=2cm, minimum width=1.5cm, minimum height=1cm] (high) {};
        \pic [pic text={High-level\\ controller},right of=sensor1, node distance=2cm] {multinode};
        
    \node [draw, rectangle, fill=gray!40, below of=high, node distance=3.2cm, minimum width=0.4\textwidth, minimum height=2cm] (env) {Environment\\[2em]};

    \node [draw, fill=red!10, rectangle, right of=high, node distance=2.5cm, minimum width=2cm, minimum height=1cm] (low1) {Low-level\\ Controllers 1};
    \node [draw, fill=blue!50, rectangle, below of=low1, node distance=1.5cm, minimum width=2cm, minimum height=1cm] (low2) {Low-level\\ Controllers 2};

    \draw [->] (sensor1) -- (high);
    \draw [->] (sensor2) -- (high);
    \draw [->] (high) -- (low1);
    \draw [->] (high) -- (low2);
    \draw [->] (low1) -| (7.8,-4) -- (env);
    \draw [->] (low2) -| (7.7,-3.5) -- (env);
    \draw [->] (env) -| (1,0) -- (sensor1);
    \draw [->] (env) -| (1.1,-1.5) --  (sensor2);
    \draw [->] (env) -| (-1,0.1) -- (driver1);
    \draw [->] (env) -| (-1,-1.5) --  (driver2);
    \draw [->] (driver1) -| (0.8,1) -| (high);
    \draw [->] (driver2) -| (0.9,0.8) -| (high);
        \node [draw, rectangle, fill=gray!40, below of=high, node distance=3.2cm, minimum width=0.4\textwidth, minimum height=2cm] (env1) {Environment\\[2em]};

    \begin{scope}[shift={(env.south west)}, scale=0.6, align=left]
        \draw[very thick, color=white, dashed] (0,2) -- (0.67\textwidth,2);
                           \node [sedan top,body color=blue!50,window color=black!80,minimum width=4cm] at (0.2\textwidth,1) {\Large 2};
                    \node [sedan top,body color=red!10,window color=black!80,minimum width=4cm] at (0.5\textwidth,1) {\Large 1};
    \end{scope}
\end{tikzpicture}
 
 \caption{A common automated driving architecture for vehicles~1 and 2. Differing components are colored in red and blue, respectively. Low-level controllers are configuration-specific, while each vehicle has an instance of the same high-level controller implementation.}
\label{fig:scheme}
\end{figure}

A typical architecture for an \ac{ad} system \cite{Widmann2000} is presented for two vehicles with different \ac{vhc}s in Fig.\,\ref{fig:scheme}. Each vehicle is equipped with different sensors and actuators, necessitating distinct low-level controllers. Both \ac{ad} vehicles are assumed to use the same software stack for the high-level controller. For longitudinal guidance, the high-level controller \ac{acc} generates an acceleration $a$, which is provided to the low-level controllers such as the engine control unit, powertrain, transmission, and brake controllers. 

In Fig.\,\ref{fig:scheme}, Vehicle~1 serves as a relevant front object for Vehicle~2. Let $h$ denote the current headway between the two vehicles. The time headway $t_{h}$ represents the duration until a collision occurs between Vehicle~1 as a front object for Vehicle~2, assuming Vehicle~1 suddenly stops while Vehicle~2 continues moving at its current speed $v$; this is expressed as $t_h=h/v$. ACC can be activated or deactivated by the driver, who also specifies a desired velocity $v_d$ and a desired time headway $t_{h_d}$. The desired time headway $t_{h_d}$ to the front object indicates the relative distance that must ultimately be maintained. For simplicity, it is assumed that the driver input interface is the same in all vehicles. The driver inputs form a parameter vector $p_t = [v_d,t_{h_d}]^T$, bounded by
\begin{align}\label{eq:params}
 P=\{p_t\in[p_{{min}},p_{{max}}]\}.
\end{align}

In set speed operation, the ego vehicle's velocity has to eventually reach the driver's desired velocity. The front object is not a concern and only the vehicle's internal actuator limitations need to be respected. In terms of portability, this can be checked using simple testing. In time gap (keep distance) operation, the sensors gather information about relevant front objects, which the ACC uses to determine the desired acceleration $a$ automatically. The ACC ensures that the headway to the front object remains above a minimum value $h_{min}$, while also keeping the current time headway above a minimum time headway $t_{h_{min}}$. Assuming a bounded velocity of the ego vehicle and the front object, this corresponds to the \ac{odd}:
\begin{align}\label{eq:collision}
 \begin{aligned}
 O=\{h,v,v_T|&h\geq h_{min}, h/v \geq t_{h_{min}},\\ &v_{min}\leq v \leq v_{max}, v_{T,min}\leq v_T \leq v_{T,max}\}.
\end{aligned}
 \end{align}

%% file: sections/framework.tex
\section{Checking functional portability}\label{sec:framework}

The goal of this work is to automatically check the safe functional portability of an ACC (denoted by the high-level controller in Fig.\,\ref{fig:scheme}), implemented in a general purpose programming language (e.g., C/C++), for different \ac{vhc}s. For that, a model of a \ac{vhc} $n$, denoted by $\sigma_n$, has to capture safety-relevant behavior. Let the set of all \ac{vhc} models be $\Sigma=\{\sigma_1,\ldots,\sigma_N\}$. Then, safe functional portability is ensured, when the controller implementation provides only controls $a$ for every relevant platform model $\sigma_n\in \Sigma$, such that the closed-loop operation always remains in \eqref{eq:collision} for all parameters \eqref{eq:params}. Testing the portability of keep distance behavior for different \ac{vhc}s is non-trivial because the \ac{odd} $O$ may vary for each hardware setup. In addition, due to, e.g., the internal actuator limitations of the vehicles, not the whole area of $O$ is safe.

First, a formal model is developed that incorporates configuration-specific effects that can compromise safe portability. The workflow depicted in Fig.\,\ref{fig:diagramm} is employed to check portability, using the framework introduced in \cite{Nenchev2024}. A safe set $S_n$ within the \ac{odd} $O_n$ is computed for each \ac{vhc} model $\sigma_n$, where every trajectory originating from this set remains contained within it at all times. To minimize excessive simulation and ensure completeness, checking portability of the high-level controller \ac{acc} for each \ac{vhc} is formulated as a containment problem with respect to the safe set $S_n$ and parameter set $P_n$. Next, we focus on modeling vehicle configurations.

\tikzstyle{traj_plan} = [rectangle, 
minimum width=3cm, 
minimum height=1cm, 
text centered, 
text width=3cm, 
draw=black, 
fill=white]

\tikzstyle{plant} = [rectangle, 
minimum width=3cm, 
minimum height=1cm, 
text centered, 
text width=3cm, 
draw=black, 
fill=white]

\tikzstyle{plan_box} = [rectangle, 
minimum width=1cm, 
minimum height=1cm, 
text centered, 
text width=1.5cm, 
draw=white, 
fill=white]

\tikzstyle{cost_calc} = [rectangle, 
minimum width=3cm, 
minimum height=1cm, 
text centered, 
text width=3cm, 
draw=black, 
fill=white]

\tikzstyle{param} = [rectangle, 
minimum width=1cm, 
minimum height=1cm, 
text centered, 
text width=1.5cm, 
draw=white, 
fill=white]

\tikzstyle{sum} = [circle, 
node distance=1cm, 
draw=black]

\tikzstyle{arrow} = [->,>=stealth]
\tikzstyle{pinstyle} = [pin edge={<-,thin,black}]

\begin{figure}[t]
\centering
\resizebox{0.5\textwidth}{!}{
    \begin{tikzpicture}[
    every node/.style={font=\sffamily\small, node distance=2cm},
    block/.style={
        rectangle, draw, align=right, minimum width=0.33\textwidth,
        minimum height=3.5cm, font=\sffamily\small
    },
    multinode/.pic={
        \node [block, \tikzpictextoptions](n) at (0,0) {\tikzpictext};
        \path [block] 
          ([shift={(.1,0)}]n.north west)
          --+(0,.1)
          -|([shift={(.1,.1)}]n.south east)
          --+(-.1,0)
        ;
    }]
        \pic [pic text={}] at (9,0) {multinode};
    \node (traj_plan) [traj_plan] {Adaptive Cruise Controller Code};
    \node (plant) [plant, right of=traj_plan, xshift=2cm] {Checker};
            \node (plan_box) [plan_box, right of=plant, xshift=1.2cm] {Safe set $S_n$};
    \node (cost_calc) [cost_calc, right of=plan_box, xshift=1cm] {Safe Set Computation Engine};
    
        \node (param) [param, below of=plan_box, yshift=0.8cm] {Parameter set $P_n$};

    \draw [arrow] ([yshift=0.1cm]traj_plan.east) -- node[above] {$u$} ([yshift=0.1cm]plant.west);

    \draw [] (plant.north) - ++(0, 0.4);
    \draw [arrow] ($(plant.north)+(0,0.4)$) -| node[left] {$x,p$}(traj_plan.north);
    \draw [thick, dashed] ($(traj_plan.west)+(-0.25,1.22)$) rectangle ($(plant.east)+(0.25,-1.22)$);
    
    \draw [] ($(plant.east)+(0.25,0)$) - ++(0.5, 0) node[anchor=north, yshift=0.7cm] {};
        \draw [arrow,thick] (cost_calc.west) -- node[above] {} (plan_box.east);
        \draw [arrow,thick] (plan_box.west) -- node[above] {} (plant.east);
        \draw [arrow,thick] (param.west) -- node[below] {} (plant.south);
     \draw [<-,thick] (cost_calc.north) -- ++(0, 0.5) node[centered, yshift=0.2cm] {\ac{odd} $O_n$};
          \draw [<-,thick] (cost_calc.south) -- ++(0, -0.5) node[centered, yshift=-0.2cm] {\ac{vhc} model $\sigma_n$};
    \end{tikzpicture}
}
 \caption{Portability checking for each \ac{vhc} $n$: based on the \ac{odd} $O_n$ and the model $\sigma_n$, the safe set $S_n$ is computed. Then, $S_n$ is used with the parameter set $P_n$ to check the safe closed-loop operation of the ACC software.}
 \label{fig:diagramm}
\end{figure}
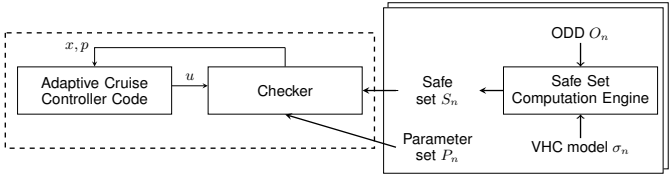

\subsection{Vehicle hardware configuration modeling}

To use model checking for functional portability, effects of the \ac{vhc} (and low-level software) that have an impact on ACC safety have to be included in a representative model. 

\subsubsection{Vehicle dynamics}\label{sec:model}

As ACC is concerned with the longitudinal motion of the vehicle, a dynamical longitudinal vehicle model is required to describe the vehicle's motion along a path. For that, force-balance equations based on Newton’s second law of motion are employed. The vehicle is assumed to move on flat terrain, and exact feedback linearization is used to compensate non-linearities from the aerodynamic drag and rolling resistances \cite{Rajamani2011}. Remaining unmodelled dynamical effects are captured by an additive bounded disturbance $w \in W, W = [w^n_{min},w^n_{max}]$ to the AD vehicle's velocity. Thus, the vehicle model is given by
\begin{align}\label{eq:cont_model}
\dot{v}=c_{n,1} a + c_{n,2}w, \dot{v_T}=a_T, \dot{h}=v_T-v,
\end{align}
where $c_{n,i}\in(0,1]$ denote configuration-specific parameters. The \ac{ad} vehicle acceleration $a$ and the object acceleration $a_T$ are bounded with $U=\{u_t = [a_t,a_{T,t}]^T\in[a_{{min}},a_{{max}}]\times[a_{T,{min}},a_{T,{max}}]\}$. As the linear model \eqref{eq:cont_model} is a simplified representation of a vehicle's behavior, additional important sensor and actuator effects are considered next.

\subsubsection{Sensor effects} 
To perceive its surroundings, a modern \ac{ad} vehicle uses a mix of cameras, LiDAR, radar, audio, ultrasound, GPS, and inertial measurement. Object detection effects are of main interest for checking functional portability of ACC software. 

Different sensors have varying abilities to detect objects based on distance and angle. Current sensor models for testing \ac{ad} systems are categorized into low-fidelity (ideal), medium-fidelity (probabilistic), and high-fidelity (physical) models \cite{Schlager2020}. Low-fidelity and some medium-fidelity models provide both object lists and perception data, while high-fidelity models only offer raw measurement data. This paper considers the former type of models, as it examines high-level ACC controllers. Each sensor setup has a maximum distance at which objects can be reliably detected. For each \ac{vhc} $n$, a specific maximum detection distance $h_{max,n}$ is estimated, which is used to define a configuration-specific object detection range. The general \ac{odd} in \eqref{eq:collision} is extended by $h\leq h_{max,n}$, yielding a configuration-specific \ac{odd} $O_n$.

The ability to classify objects also varies by sensor type. For instance, a radar may correctly identify a vehicle but fail to recognize a pedestrian in the same position. This object-dependent perception effect is considered by augmenting the parameter set \eqref{eq:params} by an additional parameter for object classification $c\in\{\text{car},\text{pedestrian}\}$, yielding a configuration-specific parameter set $P_n$.

Measurement uncertainty typically occurs within a limited range around the actual value. An additive disturbance $w$ affecting the AD vehicle's velocity $v$ has already been introduced in \eqref{eq:cont_model}. A uniform distribution is assumed to account for measurement uncertainties. Given the dynamics \eqref{eq:cont_model}, the disturbance $w$ also acts indirectly on the headway. This approach helps to evaluate the performance of different sensor setups across various vehicle configurations by treating all potential errors as equally likely. It is important to note that the formal approach presented in this paper is best used in combination with evaluations of portability based on representative real-world data.


\subsubsection{Actuator effects}

When ACC is deployed in the vehicle, there is a significant time delay in executing the input signal by the actuators. This is often caused by the delays in vehicle communication buses or by the dynamics of low-level controllers and actuators. As many AD functions are safety critical, they have to maintain a fixed cycle time $t_{cycle}$ for obtaining a new measurement from the environment through the sensors and providing an updated control $a$. If a \ac{vhc} has lower onboard computing capabilities, but the same software is to be deployed, it is possible that an updated control is produced in a following cycle. All these effects accumulate to a substantial control delay $\tau_n$ that may compromise safety in a critical situation, e.g., when braking is required instantaneously to avoid a collision. 

Considering the aforementioned effects, a combined dynamical model is developed. The overall state at the discrete time $t$ is given by $x_t=[v,v_T,h]^T$, which contains all variables relevant for the \ac{odd} \eqref{eq:collision}. The continuous differential equations \eqref{eq:cont_model} are converted into discrete time difference equations using exact discretization with an equidistant sampling time $t_s=t_{cycle}$. A zero-order hold is used at a time instant $t$ for the duration of $t_s$ for $a$ and $a_T$, which are denoted by $a_t$ and $a_{T,t}$ in the discrete time domain. Without loss of generality, the overall control delay is $\tau_{n} = k_n t_{s}$ for a given fixed worst-case $k_n\in\mathbb{N}$ for \ac{vhc} $n$. Let $a_{t-\tau_n}$ denote the control with $\tau_n$ time delay. Thus, the overall model for each \ac{vhc} $\sigma_n$ is
\begin{align}\label{eq:model_lon}
\begin{aligned}
&x_{t{+}1}{=} A x_t {+} B^a_n a_{t-\tau_n} {+}B^f a_{T,t} {+} E_n w_t, B^f{=}\begin{bmatrix}
    0&
    t_s &
    \frac{1}{2} t_s^2
  \end{bmatrix}^T,\\
&A{=}\begin{bmatrix}
    1 & 0&0 \\
    0 & 1 & 0\\
    -t_s&t_s&1\\
  \end{bmatrix}, B^a_n{=}\begin{bmatrix}
    c_{1,n} t_s \\
    0 \\
    -\frac{c_{1,n}}{2} t_s^2\\
  \end{bmatrix}, E_n{=}\begin{bmatrix}
    c_{2,n} t_s \\
    0 \\
    -\frac{c_{2,n}}{2} t_s^2\\
  \end{bmatrix}.
\end{aligned}
\end{align}

\subsection{Safe set computation} 

As described in the previous section, each \ac{vhc} $n$ may have a different 
\ac{odd} represented by a convex polytope over states and inputs, i.e., $O_n=(O\cap \{h\leq h_{max,n}\})\times U$. To check the safe operation of a high-level controller (Fig.\,\ref{fig:scheme}), control signals that keep a subset of the \ac{odd} invariant need to be identified. Note that $a_{T,t}$ is considered as an unknown disturbance for the model. The dynamical model with delay \eqref{eq:model_lon} is represented equivalently by a higher-dimensional auxiliary system without delay using the Krasovskii approach, which adds states for delayed inputs \cite{Laraba2016}. Then, the maximal Robust Controlled Invariant Set (RCIS) is computed for the discrete-time auxiliary system with ODD $O_n$ using, e.g., \cite{Holaza2023}, and the corresponding safe set $S_n$ is represented by $N_{S}$ inequalities, i.e., $S_n=\{x|A^n_x x \leq B^n_x\}, A^n_x\in\mathbb{R}^{N_{S} \times 3}, B^n_x\in\mathbb{R}^{N_{S}}$, where $A_x^n$ and $B_x^n$ are matrices defining the set.

\subsection{Checker}
A controller implementation is safe with respect to $\sigma_n$ and the \ac{odd} $O_n$ when it operates only within their safe set $S_n$. The Checker in Fig.\,\ref{fig:diagramm} is a \ac{bmc} that executes the controller code to verify that it cannot produce a control signal that leads to leaving the safe set. Since the parameters $v_d$, $t_{h_d}$ and $c$ can only take finite values, checking is done for each feasible combination thereof in parallel. 

For a state $x_t\in S_n$, there exists at least one control action $a_t$, front object acceleration $a_{T,t}$ with $[a_t,a_{T,t}]'\in O_u$ and a disturbance $w_t\in W$, such that the next state $x_{t+1}$ remains within $S_n$. Thus, the safe portability of the ACC controller is checked as follows. The continuous variables $x_t$, $a_{T,t}$ and $w_t$ are chosen by the \ac{bmc} within their admissible bounds using \texttt{assume} statements and provided to the to-be-checked controller code that produces an output $a_t$. Checking that the next state resulting from $x_t$, $a_{T,t}$, $w_t$ and $a_t$ with \eqref{eq:model_lon} is within the safe set is evaluated using an \texttt{assert} statement. The BMC also implicitly checks for implementation and security flaws such as integer overflows and pointer errors.

For a DNN-based controller, the Checker (Fig.\,\ref{fig:diagramm}) is based on a three-step approach: first, a finite number of representative states within the \ac{odd} are selected to test integration. Next, the original DNN is replaced with a simpler version and BMC is used to verify that all deployment code works correctly. Finally, a dedicated DNN verifier is used to check if the original DNN produces safe $a_t$ for a $x_t$ in the safe set and admissible values of $a_{T,t}$ and $w_t$. Safe functional portability is confirmed only if all three verification steps are successful. The interested reader is referred to \cite{Nenchev2024} for further details.

%% file: sections/case_study.tex
\section{Case study}\label{sec:experiments}
The proposed approach is applied for checking the safe portability of two adaptive cruise controllers for three \ac{vhc}s. CBMC 5.95.1 \cite{Clarke2004} is used as a bounded model checker with a timeout of 1 hour. The analysis was performed on a workstation with an AMD Ryzen Threadripper PRO 7955WX (16-Core) CPU with 128 GB RAM. 

\subsection{Vehicle hardware configurations}

All three vehicle configurations are assumed to be equipped with a typical set of AD sensors. The parameters $v_t,v_{T,t}\in[1,130]\,km/h$, $h_{min}=5\,m$, $t_{h_{min}}=0.9\,s$ are common for all configurations. In addition, the boundaries of the front object acceleration are assumed as $a_{T,min}=-1\,m/s^2$ and $a_{T,max}=0.5\,m/s^2$. Without loss of generality, a fixed desired ego velocity $v_d=130\,km/h$ and a desired time headway $t_{h_d} = 1.8\,s$ are assumed in the case study. 

\begin{figure}[t]
\centering
\begin{subfigure}{0.5\textwidth}
 \includegraphics[width=1\textwidth]{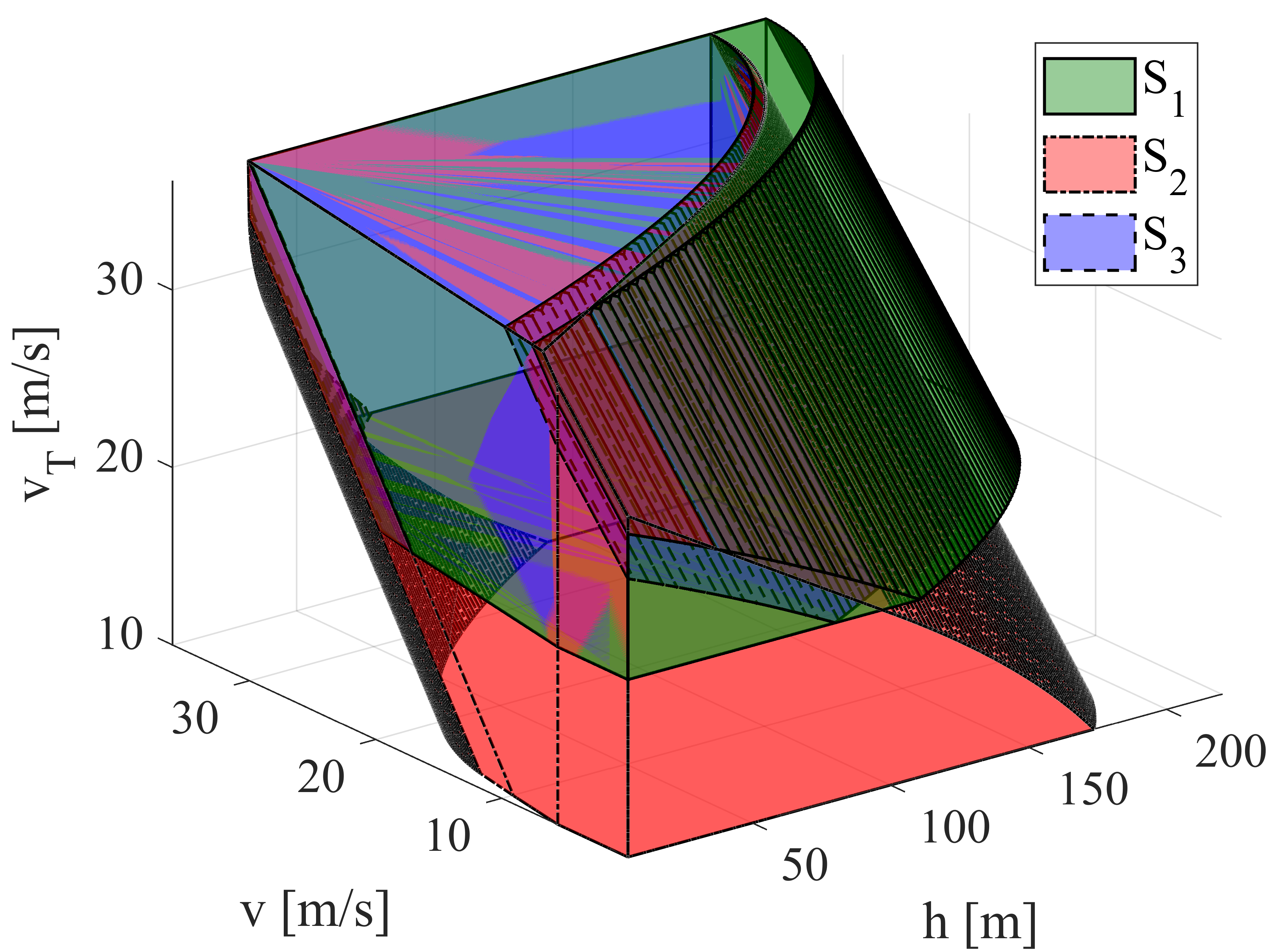}
\end{subfigure}
\hfill
\begin{subfigure}{0.5\textwidth}
    \includegraphics[width=1\textwidth]{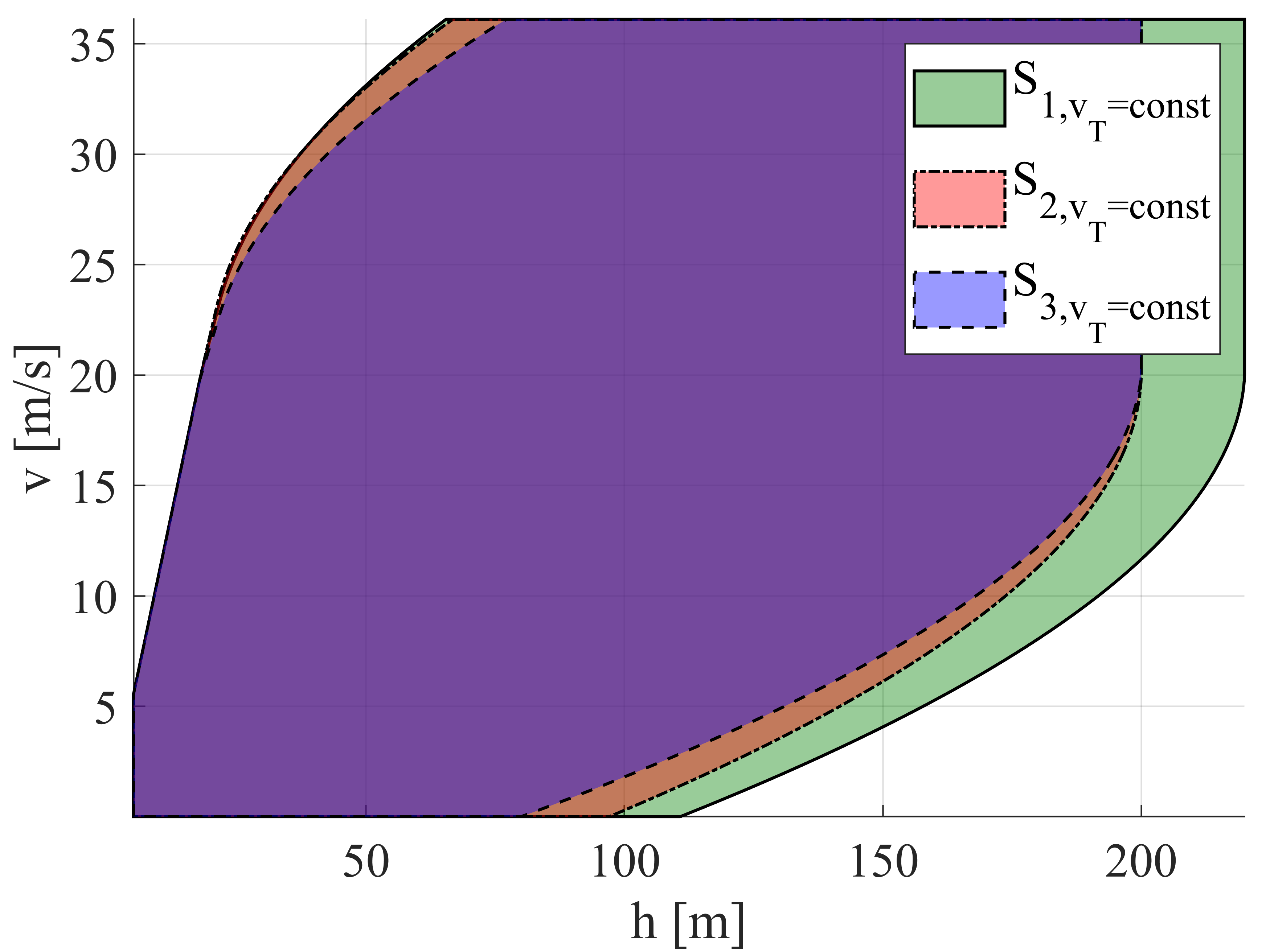}
\end{subfigure}
 \caption{Slices of the robust controlled invariant sets: 3D for $w_t=0$ and zero delayed inputs (upper plot), and 2D for additionally $v_{T}=20\,m/s$ (lower plot).}
 \label{fig:safe_set}
\end{figure}

A small set of reference maneuvers is defined according to \cite{Menzel2018} and the model parameters are tuned to reproduce the corresponding driving behavior using local optimization methods such as \cite{Wu2024}. The estimated values of configuration-specific parameters and bounds are summarized in Table~\ref{tab:params}.
\begin{table}[t]
\centering
\caption{Estimated parameter values for the vehicle hardware configuration models.}
\begin{tabular}{llrrr}  
\toprule
Parameter& Description & VHC 1& VHC 2&VHC 3\\
\midrule
   $c_{1}$&friction/drag coef& 0.95 & 0.9 & 0.85  \\
      $c_{2}$&friction/drag coef& 0.1 & 0.2 & 0.2 \\
      \midrule
         $t_{cycle} [s]$ &cycle time & 0.2 & 0.1 & 0.2  \\
   $k$ &time delay coef& 1 & 1 & 2  \\
   \midrule
   $h_{max} [m]$&max sens. range & 220   & 200   & 200     \\
   \midrule
      $a_{min} [m/s^2]$&min acceleration& -4   & -4   & -4    \\
            $a_{max} [m/s^2]$&max acceleration& 2   & 2.2   & 2.1    \\
                  $w_{min} [m/s^2]$&min disturbance& -0.05   & -0.1   & -0.1    \\
            $w_{max} [m/s^2]$&max disturbance& 0.05   & 0.1   & 0.1    \\
\bottomrule
\end{tabular}
\label{tab:params}
\end{table}
The safe sets $S_1$ of \ac{vhc}~1 and $S_2$ of \ac{vhc}~2 are computed based on the maximal RCIS of \eqref{eq:model_lon} augmented by an additional state for a one step actuation delay within their respective \ac{odd}s $O_1$ and $O_2$. Similarly, the safe set $S_3$ of \ac{vhc}~3 is obtained by using two additional states for a two-step actuation delay. The maximal number of iterations for safe set computation with \cite{Holaza2023} is 100. Under the assumption that all delayed input states and $w_t$ are zero, a 3-dimensional slice of the safe set polytopes is illustrated in Fig.\,\ref{fig:safe_set}. Additionally, a 2-dimensional slice with $v_T=20~m/s$ is presented. Note that $S_2$ has a larger volume than $S_1$ and $S_3$ mainly due to the shorter cycle time $t_{cycle}$. 

\subsection{Controller implementations}
The first considered type of ACC is based on a \ac{mpc} \cite{Naus2010}, using the model \eqref{eq:model_lon}, but not accounting for the additive disturbance $w_t$. Such controllers are commonly used in automated driving and other industrial applications. The front object is assumed to maintain $a_{T}=0$ throughout the optimization horizon with $N=5$ samples. Starting from the initial state $x_{\tilde{t}}|_{\tilde{t}=0}$, the following quadratic program is solved at each state $x_t$:
\begin{align*}
\begin{aligned}
\text{min}_{a_{\tilde{t}}} &\sum_{\tilde{t}=0}^{N} (\|v_{\tilde{t}}-\text{min}(v_d, h_{\tilde{t}}/t_{h_d})\|+\|a_{\tilde{t}}\|),\\
\text{s.t. }& \forall \tilde{t}\in[0,N],x_{t+1}=Ax_t + B_n^a a_t+B^f a_{T,t};\\
&\forall\tilde{t}\in[1,N], a_{T,\tilde{t}}=0; a_{\tilde{t}} \in U; x_{\tilde{t}} \in O;x_{\tilde{t}}|_{\tilde{t}=0}=x_t.\\
\end{aligned}
\end{align*}
The controller is implemented using the Multi-Parametric Toolbox \cite{Holaza2023}. An explicit solution with $348$ state feedback controllers was exported to C, which translates to 5521 lines of C code.

The second considered type of ACC is based on a Neural Network Controller (NNC) \cite{Zhu2020}. This approach combines imitation learning from recorded demonstrations with the optimization of a reward function that includes safety, efficiency, and comfort metrics to maximize cumulative rewards through simulations. Deep deterministic policy gradient is used to learn an actor network together with a critic network. The focus is on verifying the actor with an input $x_t$ and an output $a_t$. The actor has one hidden layer with 30 neurons, using the rectified linear unit activation function for all layers. The deployment code for this NNC consists of 1672 lines.

\begin{table}[t]
\centering
\caption{Computation times for RCIS in $[min]$ and falsification/verification in $[min] / [min]$ for controllers for different \ac{vhc}s: `-' left: checking successful, `-' right: counter-example found. 
}
\vspace*{1.5mm}
\setlength\tabcolsep{3pt}
\begin{tabular}{lccc}  
\toprule
operation & VHC~1& VHC~2&VHC~3\\
\midrule
  RCIS computation& 7.5 & 29.5 & 12.1 \\
  RCIS inequalities& 3642& 9460&4199\\
\midrule
   MPC checking& -- / 3.2 & -- / 7.0 & 5.2 / -- \\
      NNC checking   & 16.4 / --   & -- / 35.4   & 30.1 / --   \\
\bottomrule
\end{tabular}
\label{tab:results}
\end{table}

\subsection{Results} 
Table\,\ref{tab:results} shows results from the RCIS computation for the corresponding \ac{vhc}s and the portability check of the controllers. The MPC, which was designed to explicitly consider the model \eqref{eq:model_lon} and the operation set, was verified for \ac{vhc}~1 and 2, but not for \ac{vhc}~3 due to the higher control delay. Interestingly, the NNC was not verified for \ac{vhc}~1 and 3, but it was verified for \ac{vhc}~2. This suggests that the control delay plays a more important role than the maximum sensing range for the safety of this configuration. All falsifying scenarios denote an insufficient ego vehicle deceleration, while the front object decelerates with $a_{min}$ and the time headway is not large enough. 

\subsection{Discussion}
The experimental results demonstrate several benefits of our approach. The portability of controllers were either confirmed or a counter-example, which hints for necessary adaptations of the corresponding \ac{vhc} or software, was provided (Table~\ref{tab:results}). This allows an evaluation of vehicle hardware effects on AD safe portability, and the possible trade-offs that can be made between the available hardware, the desired \ac{odd} and the implemented software. For example, the confirmed portability of NNC for \ac{vhc}~2 and the discovered counter-example for \ac{vhc}~1 suggest that safety can be maintained with a smaller control delay even with a lower maximum sensing range. Thus, the method can be used to gather feedback for potential \ac{vhc}s allowing safe operation with a lower cost or a larger \ac{odd}. The proposed portability check is suitable for time-invariant and deterministic controllers. Note that the obtained \ac{vhc} safe sets can be used as customized supervisors for legacy \ac{acc}. Even though ACC was in focus, the presented check is not limited to the considered \ac{odd} and model \eqref{eq:model_lon}. A similar approach can be adopted for other cyber-physical systems. With automated checking possible within minutes, the solution can be integrated into a continuous integration process.

Despite the promising results in the case study, several limitations must be noted. The portability check provides either a compliance proof or a counter-example based on the employed model and specification. The safe portability check is complete across the \ac{odd}; however, it is based on a formal model that, in general, does not describe all behaviors of the actual system. Real-world data may not cover the whole ODD completely, potentially leading to gaps in evaluation. Consequently, the presented approach should be viewed as complementary to assessments using real-world data. An alternative remedy might be to complement the formal model with statistical validation methods in addition to traditional discrepancy analysis techniques \cite{Dona2022}. Further, defining the operation and parameter sets for model checking becomes more challenging with an increasing model and specification complexity. Finally, high fidelity models may render RCIS computation infeasible or automated checking intractable. Specifically, larger actuation delays greatly increase the complexity of RCIS computation. In addition, large disturbance sets may lead to an empty RCIS.

%% file: sections/conclusions.tex
\section{Conclusions}\label{sec:conclusions}
A portability checking approach based on formal modeling of safety-critical aspects of vehicle hardware configurations was proposed for adaptive cruise control software. By computing a safe set for these models, the portability of the controller implementation was automatically checked using an existing verification framework. This approach was successfully applied to both a conventionally implemented controller and a neural-network-based controller. The results highlight the potential of formal methods to streamline the adaptation process for automated driving systems, providing a scalable and reliable solution to what is typically a labor-intensive and error-prone integration process.

While the focus was on high-level controller code, low-level software (such as drivers and OS routines) has not been considered. Since the latter are often handwritten for each hardware and may cause additional safety flaws, these should be considered for portability in the future. Further, modeling additional effects from sensors and actuators, as well as other automotive controllers will be studied.